\documentclass[superscriptaddress,twocolumn,aps,prl,showpacs,preprintnumbers,amsmath,amssymb]{revtex4}
\usepackage{epsfig}
\usepackage{amssymb,amsfonts}

\begin{document}
\def\mydag{^{\vphantom{\dagger}}}
\title{Dynamics of quantum light in integrated nonlinear waveguide arrays and generation of robust continuous variable entanglement  }

\author{Amit Rai}
\affiliation{Centre for Quantum Technologies, National University of Singapore, 3 Science drive 2, Singapore 117543.}

\author{Dimitris G. Angelakis }
\affiliation{Centre for Quantum Technologies, National University of Singapore, 3 Science drive 2, Singapore 117543.}

\affiliation{Science Department, Technical University of Crete, Chania, Crete, Greece, 73100}

\date{\today}

\begin{abstract}

\noindent We study a class of nonlinear waveguide arrays where the waveguides are endowed with $\chi^{(2)}$ nonlinearity and are coupled through the evanescent overlap of the guided modes. We study both the stimulated and spontaneous process in the array and  show the viability of such an array as a platform for generating both bipartite and tripartite continuous variable entanglement on demand. 
 We explicitly address the effect of realistic losses on the entanglement  produced, briefly discuss the possible types  of nonlinear materials that could be used, and suggest solutions to the possible phase matching issues in the waveguides. The simultaneous generation and manipulation of the light on a single waveguide chip circumvents the usual bandwidth problems associated with the use of external bulky optical elements and makes this avenue promising for further investigation
 
 \end{abstract}


\maketitle

\noindent

\textit{Introduction:} The generation and manipulation of light are at the heart of optical physics and photonics technologies. In particular, waveguides offer highly flexible tools for manipulating and processing light over short distances and have thus found applications in diverse areas of research \cite{Saleh}. For example, in optics they form the building block of a more complex structure commonly known as waveguide arrays \cite{Christodoulides}. The possibility to manipulate various interactions by design makes these arrays an experimentally accessible tool for studying a variety of effects from a large number of fields of physics. Another salient feature of this system is the possibility to control the exact initial conditions for the light propagating inside the array. Moreover, the decoherence
rate in this system is very low even for longer propagation distances \cite{Quantumwalk}.
Some of these studies include realization of condensed-matter-like effects  \cite{Anderson, Lahini, BLOCHOSCILLATION0,BLOCHOSCILLATION1,BLOCHOSCILLATION2,BLOCHOSCILLATION3, NOONSTATE, WSTATE}, quantum random walks \cite{Quantumwalk,Bromberg11}, and the quantum Zeno effect \cite{QUANTUMZENO}. Quite recently, it has also been shown that one can tailor the dispersion relation for the waveguides to study the famous Bose-Hubbard Hamiltonian in a classical setup \cite{BOSEHUBBARD}. Furthermore, waveguides along with single photons have been shown to form basic units for quantum network architectures \cite{OBRIENN}.

The above discussion clearly shows that the waveguide arrays are quite suitable for investigating a variety of physical effects and can also be used to effectively manipulate various quantum states of light. The question naturally arises as to whether such structures can also be utilized as a flexible tool to both generate and manipulate the quantum states of light at the same time. Here we explore the issue by studying an experimentally accessible model for the waveguide arrays where the waveguides are endowed with $\chi^{(2)}$ nonlinearity and are coupled through the evanescent overlap of the guided modes.

As a relevant application of our study, we choose to study the generation of continuos variable entanglement using the waveguide structure in an integrated manner on a single waveguide chip \cite{loock12}. We note that a detailed discussion about the entanglement in continuous variable systems can be found in Ref. \cite{Adesso}. Conventionally, the continuous wave entangled light has been generated from optical parametric process inside an optical cavity. However, the use of optical cavities severely limits the bandwidth of the entangled beam generated in such processes. It would thus be highly desirable to avoid the use of optical cavities. The use of optical waveguides where the transverse field confinement results in an increase of the nonlinear efficiency can makeup for the buildup of a reasonably high finesse cavity.

\begin{figure}
\begin{center}
\includegraphics[scale = 0.6]{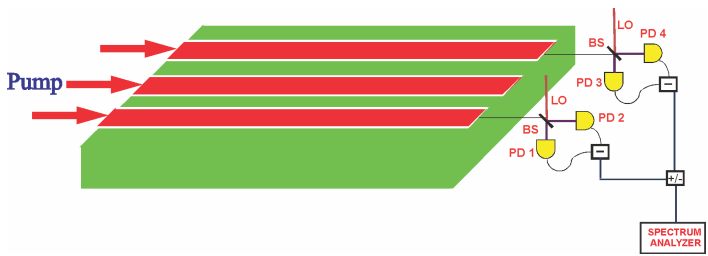}
\caption{\label{1} Schematic of a waveguide array with $\chi^{(2)}$ nonlinearity. The correlation function $M {(j,k)}$ can be measured by feeding the waveguide modes into the homodyne detectors. The signals from the homodyne detectors are subtracted or summed electronically
and the variance is measured by a spectrum analyzer. The resulting values of the variances can be used to calculate the correlation function $M {(j,k)}$. The elements are: symmetric beam splitters BS, local oscillator LO, photodiode PD.}
\end{center}
\end{figure}

\textit{System Description :} We consider a waveguide system which contains an array of $N$ identical wave\-guides endowed with $\chi^{(2)}$ nonlinearity. We assume that each of these wave\-guides are pumped through a coherent light (as shown in Fig.~\ref{1}). The waveguide arrays studied here can be implemented for example by using the periodically poled lithium niobate waveguides employed in the work of Yoshino et al. \cite{Yoshino}. The coupling between the wave\-guide is achieved by the evanescent overlap of the guided modes \cite {Iwanow}. Further, we assume that the pump field is in a strong coherent classical field, which is strong enough to remain undepleted of photons over the entire length of the waveguide. The field operators evolve according to the Heisenberg equations given by:

\begin{eqnarray}
  \dot{a}_j\mydag= -2 i g_{j}
a_j^\dagger +i J_{j} (a_{j-1}\mydag
+a_{j+1}\mydag)~,
\end{eqnarray}

\noindent where $a_{j}\mydag$($a^\dagger_{j}$) refer to the bosonic
creation (annihilation) operator for the field in the $j^{th}$
waveguide. The nonlinear coupling parameter $g_{j}$ depends on strength
of the pump laser and the nonlinear susceptibility of the waveguide medium. The linear coupling
parameter $J_{j}$ represents the rate at which the photons are
transferred to the neighboring wave\-guides. Note that we have ignored the evanescent coupling between the pump beams because the coupling coefficient for the pump beam would generally be much smaller compared to the signal ones, due to the weaker overlap between the waveguide modes at
higher frequencies \cite{Iwanow}.

\noindent The evolution of the photon annihilation operators $a_{j}\mydag$ of output modes can be written as:

\begin{eqnarray}
 a_{j}(t)  =  \sum_{k} a_{k}(0) \hspace{0.05cm} A_{j,k}+\sum_{k} a_{k}^\dagger(0) \hspace{0.05cm} B_{j,k}~,
\end{eqnarray}

\noindent where the summation is over all possible input modes. Moreover, $A_ {j,k}$ and $B_ {j,k}$ denotes the complex matrix element which depends on the coupling parameter $g$ and $J_{j}$. Also, the function $A_ {j,k}$ and $B_ {j,k}$ depends on the propagation distance, $z$, over which the light distribution evolves. We note that Eq. 2 signifies the simultaneous generation and manipulation of light in waveguide arrays. Pump photons are converted to the signal photons within the individual waveguides whereas the evanescent coupling between the signal modes leads to the linear spreading of the light across the waveguide array. 

\textit{Dynamics}: Before moving to the main part,  we first analyze the mean intensities behaviour at the output which is an obvious measurable quantity in any similar experiment. The input to the waveguide system can either be in the form of a separable or entangled state. 
For the case when the signal and idler modes are initially in the  vacuum states, the intensity will be  given by $I_{j} (t)= \sum_{l=1}^{N} ( |B_{j,l}|)^2$. In this case, the pump mode is spontaneously converted to the signal modes within the individual waveguides. Further, because of the linear coupling between the inter-waveguide modes the light spreads across the array. We note that in this case the light generated at the output of the waveguide arises solely because of the vacuum fluctuations. For the obvious most interesting case  of a coherent light  $|\alpha \rangle$ fed into the $m$th waveguide, the intensity evolution among the waveguide sites can be written as
\begin{eqnarray}
 I_{j} (t)= (|\alpha|)^2 (|A_{j,m}|^2 + |B_{j,m}|^2) +\\ \nonumber (\alpha^2 A_{j,m} B_{j,m} ^*+ h.c. )+  \sum_{l=1}^{N} ( |B_{j,l}|)^2
 \end{eqnarray}
 
\begin{figure}
\begin{center}
\includegraphics[scale = 0.52]{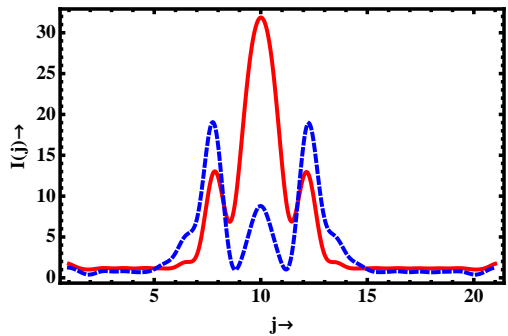}
\caption{\label{21} The intensity $I_{j} (t)$ as a
function of waveguide index $j$ for the case of $21$ waveguides. The
coupling parameters are chosen such as $g/J=1/2$ (red solid line) and $g/J=1/3$ (blue dashed line). The input to the waveguide is in $j=10$ waveguide. The input is a coherent state with amplitude $\alpha=5$. We are looking at the output intensity at $t=2.5$.}
\end{center}
\end{figure}

It is interesting here to briefly compare the quantum walk of coherent light in linear waveguide arrays as analyzed for example in \cite{Quantumwalk}, with the arrays discussed here. In the linear case, we just have the contribution from the first term for $I(t)$ and we see the ballistic propagation of the coherent light across the waveguide arrays. The quantum walk of coherent light in our system shows a very different behavior depending on the ratio of $g/J$. For higher value of $g/J$, we found that the input light is amplified and remains localized in the input waveguide. For lower value of  $g/J$, the input light spreads across the array. This is schematically shown in Fig.~\ref{21}. We note here the difference with all previous works who essentially correspond to the transfer of the quantum light while the input light itself is generated outside the waveguide system by using bulk optical elements \cite{Bromberg11, Peruzzo}. Such schemes involving bulk optical elements suffer from severe limitations as far as stability and physical size are concerned and may also introduce quantum decoherence.

\textit{Generating continuos variable entanglement }: As a relevant application to our system, we investigate the possibility of generating continuous variable entanglement in an integrated manner. We
define the quadrature operators for the $j^{th}$ waveguide given by
$q_{j} \equiv (a_{j}\mydag e^{-i \phi}+a^\dagger_{j}e^{i
\phi})/\sqrt{2}$ and $ p_{j} \equiv (a_{j}\mydag e^{-i
\phi}-a^\dagger_{j}e^{i \phi})/\sqrt{2} \hspace{0.05 cm} i$. We use
the criterion of Ref. \cite{duan} for studying the entanglement between the waveguide modes and define the correlation between two waveguide modes
as $ M(j,k) =  \langle a_j^\dagger a_{j}\mydag\rangle+ \langle a_k^\dagger a_{k}\mydag\rangle+ e^{-2 i \phi} \langle a_j\mydag a_{k}\mydag\rangle+e^{2 i \phi} \langle a_j^\dagger
a_k^\dagger\rangle$. The correlation function $M(j,k)$, our entanglement measure, can be probed by using the homodyne detectors scanning across the waveguide array output as shown in Fig.~\ref{1}. We note that a similar correlation measurement has been carried out in the work of Peruzzo et al. \cite{Peruzzo} although we emphasize that the detection process in our case is fundamentally different  the one reported in that work. In the experiment by Peruzzo et al. the output from the waveguide modes are detected with single photon detectors. On the other hand, in our case the waveguide modes at the output are mixed with strong local oscillator as shown in Fig. 1. Since the local oscillator amplifies the waveguide modes one can use high efficiency detectors that work only with strong signals. Further, note that we have written $M(j,k)$ in terms of photon annihilation and creation operators of the waveguide modes but in the actual homodyne measurement one measure the quadrature observable which can then be used to calculate the correlation function $M(j,k)$. The negativity of $M$ is a sufficient condition for entanglement. We first discuss our analytical results for the bipartite case of two
wave\-guides which read
\begin{eqnarray}
M(1,2) = 4 \sin^2(\Omega t)g(2 g-J)/{\Omega^2}
\end{eqnarray}
 where $\Omega \equiv (J^2-4 g^2)^{1 / 2}$. Clearly
$M(1,2)<0$ for $J >2g$ and thus entanglement occurs. For the case of five waveguides, we show our
numerical results in Fig. 3. The negative values of $M
(i,j)$ in Fig. 3 clearly demonstrate the entanglement
between the inter-waveguide modes. 

The system of waveguides we propose can also act as a compact source of tripartite entangled light. 
van Loock and Furusawa \cite{loock14}, have shown that for a fully inseparable three-mode system, it is sufficient to
measure: $V_{i,j,k}  = V (X_i-{(X_j+X_k)}/\sqrt{2})+V (Y_i+(Y_j+Y_k)/\sqrt{2})< 4~,$ where the mode indices $i$, $j$, $k$ are all different, to demonstrate the inseparability. The
quadrature operators for each mode are defined as follows: $
X_{i} = a_{i}\mydag e^{-i \phi} + a_i^\dagger e^{i \phi}, Y_{i} = -i (a_{i}\mydag e^{-i \phi}-a_i^\dagger e^{i \phi})$. In our case, we choose $\phi=\pi/2$ and study the evolution of $V_{2,1,3} $ for $g/J=2/3$ for the case of three waveguides in Fig. 4 and show that the conditions 
$V_{2,1,3} < 4$ is clearly satisfied meaning  the three waveguide modes are entangled.

\begin{figure}
\begin{center}
\includegraphics[scale = 0.52]{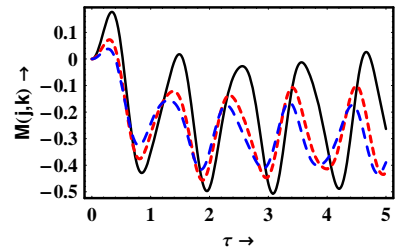}
\caption{\label{3} The correlation function $M(j,k)$ as a
function of $\tau$ ($\tau \equiv J \hspace{0.02cm}t/\pi $) for the
case of five waveguides. The
coupling parameters are chosen such as (a) $g/J=1/5$ (Black) (b) $g/J=1/7$ (Red dashed line) and (c) $g/J=1/9$
(Blue).}
\end{center}
\end{figure}

\begin{figure}
\begin{center}
\includegraphics[scale = 0.52]{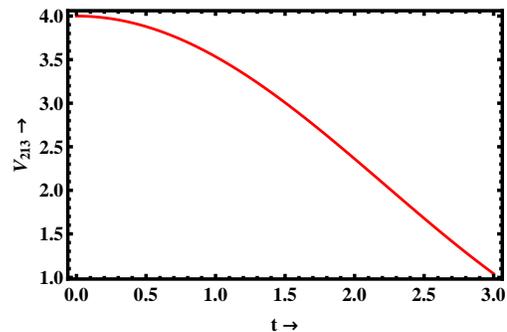}
\caption{\label{5} The correlation function $V_{2,1,3}$ as a
function of $t $ for the
case of three waveguides.The coupling parameters are chosen such as $g/J=2/3$.}
\end{center}
\end{figure}

\textit{Losses and feasibility}: Loss is the greatest challenge facing the implementation of integrated photonic technologies and is inevitable in the real world systems. Thus an immediate
question of interest would be how does this loss affects the entanglement in the waveguide
modes ? It is well known that entanglement is quite susceptible to decoherence \cite{Zurek} and thus
the above question is quite relevant in context to quantum information processing using
waveguides. Since the two wave\-guides are identical, we have taken the loss rate of both the modes to be the same. We can model the loss in
wave\-guides in the framework of system-reservoir interaction well
known in quantum optics and is given by,

\begin{eqnarray}
\label{2} \mathcal{L}\rho & =
&-\frac{\gamma}{2}(\hat{a}^{\dagger}\hat{a}\rho-2\hat{a}
\rho\hat{a}^{\dagger}+\rho\hat{a}^{\dagger}\hat{a})\nonumber\\
& & -\frac{\gamma}{2}(\hat{b}^{\dagger}\hat{b}\rho-2\hat{b}
\rho\hat{b}^{\dagger}+\rho\hat{b}^{\dagger}\hat{b})~,
\end{eqnarray}

\noindent where $\rho$ is the density operator corresponding to the system
consisting of fields in the modes a and b. The dynamical evolution
of any measurable $\langle O \rangle $ in the coupled wave\-guide
system is then governed by the quantum-Louiville equation of motion
given by $\label{3} \dot{\rho} = -\frac{i}{\hbar}[H, \rho]+\mathcal{L}\rho
$ where $\langle \dot{O} \rangle = {\rm Tr}\{O\dot{\rho}\}$, the
commutator gives the unitary time evolution of the system under the
influence of coupling and the last term account for the loss.

\begin{figure}
\begin{center}
\includegraphics[scale = 0.52]{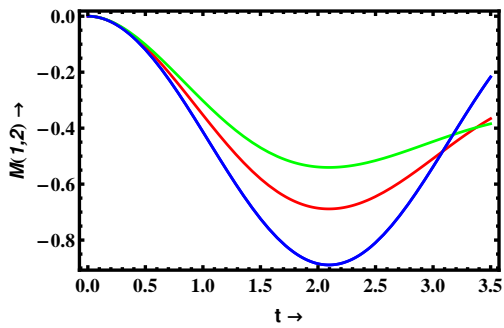}
\caption{\label{4} The correlation function $M(1,2)$ as a
function of t for the case of two waveguides in the presence of loss. The
coupling parameters are chosen such as $g/J=2/5$. The loss $\gamma$
is taken to be (a) $\gamma=0$ (Blue) (b) $\gamma/J=1/5$ (Red) and (c) $\gamma/J=2/5$
(Green).}
\end{center}
\end{figure}

\noindent We use the above equation for the density operator to evaluate the correlation $M(1,2)$ given in the presence of loss. We also set $\phi=0$ in the equation for $M(j,k)$. The results are shown in Fig.~\ref{4}. We see there that even for reasonable large values of the loss to coupling ratio the waveguides will still be entangled. Although none has ever attempted such an integrated approach for the simultaneous generation and manipulation of nonclassical light,  the experiments by Iwanow et al. \cite{Iwanow} investigate  the process of second harmonic generation in the waveguide arrays of periodically poled
$LiNbO_3$ which feature a quadratic nonlinearity. We
note that the second harmonic generation is the reverse of the down conversion process. 
Low loss waveguides with losses of $0.2 $ dB/cm for the fundamental-wave and $0.4$ dB/cm for its second harmonic were fabricated in these experiments using the technique of Quasi-Phase-Matching (QPM) \cite{Armstrong}. Because of their lower values for the loss such arrays are ideally suited for the experimental implementation of our proposal. Moreover, integrated optical circuits based on $LiNbO_3$ substrates is now very well established and a great variety of devices
based on this technology can be integrated on the single chip. For instance,
the lithium niobate waveguides have being extensively employed for applications in electro-optic switching and phase
modulators. 

Except the losses, for a successful implementation one would also need to ensure the  downconversion process to be efficient which
the usual phase matching condition must be satisfied \cite{Saleh, boyd}. The phase-matching in waveguide structures can
be achieved using the Quasi-Phase-Matching in combination with
ferroelectric materials. For example, in addition to an enhanced nonlinear efficiency due to tight confinement of the
interacting waves inside the waveguides, the QPM will allow working
with the highest non-linear coefficient of lithium niobate
($d_{33}  \approx 30 pm/V$) which is much larger
than that ($d_{31}$) commonly used in birefringent phase matching
in bulk crystals. Other materials which can be used as a nonlinear medium inside optical waveguides include periodically poled potassium titanyl phosphate (PPKTP),
quasi-phase-matched $LiTaO_{3}$, and periodically
poled stoichiometric $LiTaO_{3}$.
Recent experiments have shown that with the advancement in waveguide
fabrication techniques and improvement in the detection schemes
it is possible to achieve better than $-$$5$ dB of pulsed traveling-wave squeezing in
MgO-doped periodically poled $LiNbO_3$ (PPLN)
waveguides \cite{Eto1, Eto2}.

\noindent  \textit{Conclusion}: In conclusion, we considered the waveguide arrays with $\chi^2$ nonlinearity and studied the generation and manipulation of nonclassical light in such a system. We also investigated the possibility of generating broadband continuous variable entanglement in the waveguide structure. In our study, we propose an integrated approach towards continuous variable entanglement based on integrated waveguide quantum circuits which are compact and
relatively more stable. In addition, the use of waveguide structure also eliminates the need for the optical path alignment which is required in the bulk optical systems. The possibility to generate broadband entanglement from waveguides
could also help to avoid the use of optical cavities from many key experiments in the area of broadband continuous variable quantum information processing. We can further investigate the evolution of the Wigner functions
in the waveguide arrays which is especially interesting in the case of stimulated inputs. Further, in our study we have assumed the coupling parameter $J_{j}$ to be constant but one can also make the value of $J_{j}$ random \cite{DIAGONALDISORER} which is left for a future contribution.


\begin{thebibliography}{999}

\bibitem{Saleh} B. E. A. Saleh and M. C. Teich, Fundamentals of Photonics, 2nd Edition, (Wiley, New York 2007).

\bibitem{Christodoulides}D. N. Christodoulides, et al., Nature (London) {\bf 424}, 817 (2003).

\bibitem{Quantumwalk} H. B. Perets, et al.,
Phys. Rev. Lett. {\bf 100}, 170506 (2008).

\bibitem{BLOCHOSCILLATION0} F. Bloch, Z. Phys. {\bf 52}, 555 (1928).

\bibitem{BLOCHOSCILLATION1} U. Peschel, T. Pertsch, and F. Lederer, Opt. Lett. {\bf 23}, 1701 (1998).

\bibitem{BLOCHOSCILLATION2} T. Pertsch, et al., Phys. Rev.
Lett. {\bf 83}, 4752 (1999).

\bibitem{BLOCHOSCILLATION3} R. Morandotti, et al., Phys. Rev.
Lett. {\bf 83}, 4756 (1999).


\bibitem{NOONSTATE}  Y. Bromberg, Y. Lahini, and Y. Silberberg,
Phys. Rev. Lett. {\bf 105}, 263604 (2010).

\bibitem{WSTATE} A. Rai and G. S. Agarwal, Phys. Rev. A {\bf 79}, 053849 (2009).

\bibitem{Anderson}  P.W. Anderson, Phys. Rev. {\bf 109}, 1492 (1958).

\bibitem{Lahini}  Y. Lahini, Y. Bromberg, D. N. Christodoulides, and Y. Silberberg, Phys. Rev.
Lett. {\bf 105}, 163905 (2010).

\bibitem{Bromberg11} Y. Bromberg, Y. Lahini, R. Morandotti, and Y. Silberberg, Phys. Rev. Lett. {\bf 102}, 253904 (2009).

\bibitem{QUANTUMZENO} S. Longhi, Phys. Rev. Lett. {\bf 97}, 110402 (2006).


\bibitem{BOSEHUBBARD} S. Longhi, J. Phys. B: At. Mol. Opt. Phys. {\bf 44}, 051001 (2011).


\bibitem{OBRIENN} A. Politi, et al.,
Science {\bf 320}, 646 (2008); A. Politi, J. C. F. Matthews, and J. L. OÕBrien,
\textit{ibid}. {\bf 325}, 1221 (2009); J. L. O'Brien, A. Furusawa, and J. Vuckovic, Nature Photonics {\bf 3}, 687 (2009).


\bibitem{loock12}
S. L. Braunstein and Peter van Loock, \textit{Reviews of Modern
Physics} {\bf 77}, 513 (2005).

\bibitem{Adesso} S. L. Braunstein and A. K. Pati, Quantum Information with Continuous Variables, (Kluwer Academic Publisher, The Netherland, 2003); J. Eisert  and M. B. Plenio, Int. J. Quantum Inform. {\bf 1}, 479 (2003);  
G. Adesso and F. Illuminati, J. Phys. A: Math. Theor. {\bf 40}, 7821 (2007); U. L. Andersen, G. Leuchs, and C. Silberhorn, Laser \& Photonics Reviews {\bf 4}, 337 (2010). 

\bibitem{Yoshino} K. Yoshino, T. Aoki, and A. Furusawa, Appl. Phys. Lett. {\bf 90}, 041111 (2007).

\bibitem{Iwanow} R. Iwanow, R. Schiek, G. Stegeman, T. Pertsch, F. Lederer, Y. Min,
and W. Sohler, Phys. Rev. Lett. {\bf93}, 113902 (2004); R. Iwanow, R. Schiek, G. I. Stegeman, T. Pertsch, F. Lederer,
Y. Min, and W. Sohler, Opt. Lett. {\bf 30}, 1033 (2005); R. Iwanow, R. Schiek, G.I. Stegeman, T. Pertsch, F. Lederer, Y. Min and W.
Sohler, Opto-Electron. Rev. {\bf 13}, 113 (2005).


\bibitem{Peruzzo} A. Peruzzo, et al., Science {\bf 329}, 1500 (2010).


\bibitem{duan} L.-M. Duan, G. Giedke, J. I. Cirac, and P. Zoller, Phys.
Rev. Lett. {\bf84}, 2722 (2000); R. Simon, Phys. Rev. Lett. {\bf 84}, 2726 (2000).


\bibitem{loock14} P. van Loock and A. Furusawa, Phys. Rev. A {\bf 67}, 052315 (2003).


\bibitem{Zurek}  W. H. Zurek, Rev. Mod. Phys. {\bf 75}, 715 (2003).


\bibitem{Armstrong} J. A. Armstrong, N. Bloembergen, J. Ducuing, and P. S. Pershan, Physical Review {\bf 127},
1918 (1962).

\bibitem{boyd} R. W. Boyd, Nonlinear Optics, 3rd Edition, (Academic Press, 2008); Y. R. Shen, The Principles of Nonlinear Optics, (Wiley-Interscience, 2002).


\bibitem{Eto1} Y. Eto, T. Tajima, Y. Zhang, and T. Hirano, Opt.
Express {\bf 16}, 10650 (2008).


\bibitem{Eto2}   Y. Eto, et al., Opt. Lett. {\bf 36}, 4653 (2011).

\bibitem{DIAGONALDISORER} L. Martin, et al., Opt. Express  {\bf 19}, 13636 (2011).

 
\end{thebibliography}
\end{document}